\documentclass[aps,prd,twocolumn,nofootinbib]{revtex4}
\usepackage{graphicx}
\usepackage{graphics}
\usepackage{bbm}
\usepackage{longtable}
\usepackage{amssymb}
\usepackage{amsmath}

\begin{document}


\title{Dissipative Effect in Long Baseline Neutrino Experiments}





\author{Roberto L. N. Oliveira}
\email{robertol@ifi.unicamp.br}
\affiliation{Instituto de F\'\i sica Gleb Wataghin Universidade Estadual de Campinas, UNICAMP 13083-970, Campinas, S\~ao Paulo, Brasil} 

\date{\today}

\begin{abstract}

The propagation of neutrinos in long baselines experiments may be influenced by dissipation effects. Using Lindblad Master Equation we evolve neutrinos taking into account these dissipative effects. The MSW and the dissipative effects may change the probabilities behavior. In this work, we show and explain how the behavior of the probabilities can change due to the decoherence and relaxation effects acting individually with the MSW effect. A new exotic peak appears in this case and we show the difference between the decoherence and relaxation effects in the appearance of this peak. We also adapt the usual approximate expression for survival and appearance probabilities with all possible decoherence effects. We suppose the baseline of DUNE and show how each decoherence parameters change the probabilities analyzing the possible modification using numeric and analytic approach.

\end{abstract}


\maketitle
\section{Introduction}

In the near future there will be new long baselines experiments ~\cite{lbne, lbno, hiperk} to test the standard pattern of neutrino oscillation as never before. These experiments will have   great sensitivity to determine the parameters that describe the standard oscillation pattern. Through them will be possible investigate the open questions, like CP-violation, mass hierarchy and octant problem ~\cite{T2Kq13, MINOSq13,theoq13,CHOOZq13, bayq13,renoq13,masshi1,masshihiroshi,octant23a, octant23b}, and maybe new phenomena.  

As it is well known, the quantum mechanics explains how the neutrinos are able to change your flavors during their propagation. The success of the model opened opportunity for testing other kinds of effects in oscillation experiments ~\cite{sterile, vep, nonstand,extdimen,decay, workneut3, lis,mossbauer}. So, considering that the precision level that the next generation of experiments will achieve, the limit on each usual neutrino oscillation parameter will be very stringent and the space for new physics may also be lessened.

In this work, we will use the Lindblad Master Equation to evolve the neutrinos~\cite{lin,gor}. This equation is used when a physical system is considered open to interact with a quantum environment that is treated as a reservoir ~\cite{len, pet}. Due to interaction between the subsystem of interest and the environment, the quantum behavior of the subsystem of interest may change considerably during its quantum evolution. In the present case the neutrinos are our subsystem of interest while the current hypotheses of the dissipative sources are quantum foam or quantum gravity to neutrino propagation in matter or vacuum \cite{lis, ell, gab, gab1,ben1}, and there is  a phenomenological theory to matter fluctuation as dissipative source \cite{ben2,bur} including a microscopic model in this case, as we can find in Ref. \cite{bur}. The quantum evolution through Lindblad Master Equation is non-unitary and it adds in the evolution the possibility of occurring dissipative effects like decoherence, relaxation and other. The quantum evolution through Lindblad Master Equation is non-unitary and it adds in the evolution the possibility of occurring dissipative effects like decoherence, relaxation and other.   

In neutrino oscillation, the decoherence effect has been studied more than the other dissipative effects and the most part of these investigations, the neutrino propagation was in vacuum ~\cite{lis, ben, mmy, workneut3, fun1, dea}. However, considering the next generation of long baseline experiments, the matter interaction will be important and we will study some aspects of this case.

We will also probe the relaxation effect that is much less studied in the literature even when the propagation is in vacuum. This effect are not important for terrestrial long baseline and only the solar neutrinos are responsible for putting the most stringent bound on relaxation effect \cite{workneut4}. However, in the case of the propagation in constant matter, we are able to verify in another way the difference between the decoherence and relaxation effects. So, we will start with two neutrino approximation and will point out which the difference between the relaxation and decoherence effects when the MSW effects is present. In this case, the survival and appearance probabilities present new exotic peaks due to the presence of dissipative effect and MSW effect. This behavior will be analyzed and we will show how the differences of these effects become evident in neutrino propagation.  

As the $\theta_{13}$ value is not null, the three neutrino families will have to be taken into account in the new generation of long baseline experiments~\cite{lbno, lbne, hiperk} as well as the MSW effect \cite{kim,moh}. In this study, we will present a phenomenological approach where the most effective quantum dissipative operator will be defined to the case of three neutrino oscillations. Then, we will evolve the neutrinos in an exact and analytical scheme considering the baseline of DUNE~\cite{lbne}.

In analytical case, there is a study that can be found in Ref.~\cite{freu} where the author considered the approximation $\Delta m^{2}_{12}<<\Delta m^{2}_{31}$ to obtain analytical expressions for the oscillation probabilities in terms of the effective mixing angles. We will adapt these results to the case where the decoherence effects are taken into account and a long and short analytical expressions for the survival and appearance probabilities will be introduced. Besides, we will include a discussion about the range of valid of these new probabilities. 

We will show the behavior of these two versions of analytical probabilities in comparison with the exact approach. The exact and long analytical expression for the probabilities present the same new peak introduced due to the coupling of the decoherence and MSW effects, while the short analytical expression for the probability due to the decoherence effect is divergent. Besides, we will point out using the exact approach how each decoherence parameter changes the survival and appearance probabilities considering the baseline of DUNE.

Furthermore, considering short version of the analytical probabilities for DUNE, we are able to understand how the behavior of the survival and appearance probabilities are influenced by the decoherence effects, because the short expression is an acceptable approximation of the exact probabilities in the range energy important in DUNE. 

Then, we conclude with a simple study about CP-violation where we show how the CP-violation phenomena may be changed depending on the magnitude of the decoherence effects.


\section{Quantum Dissipators}

We introduce a dissipative formalism that may be used in long baseline experiments of the next generation like DUNE \cite{lbne}. In this context, we will consider the neutrino as an open quantum system and its propagation will be made using  the Lindblad Master Equation \cite{len, pet, joo,gor}.

This evolution equation adds many dissipative effects being the decoherence and relaxation effects the most effective of them. The physical meaning of these two effects can be found in Ref. \cite{workneut4}.

The Lindblad Master Equation is usually written as
\begin{equation}
\frac{d\rho_{\nu}(t)}{dt}=-i[H_{S},\rho_{\nu}(t)]+D[\rho_{\nu}(t)]\,,
\label{i}
\end{equation}  
where $\rho_{\nu}$ is the interesting subsystem state, $D[\bullet]$ comes from the partial trace over the environment states in the evolution equation of the global system \cite{len} and it must have energy dimension being defined as
\begin{equation}
D[\rho_{\nu}(t)]=\frac{1}{2}\sum_{k=1}^{N^{2}-1}\Big[V_{k},\rho_{\nu} V_{k}^{\dag}\Big]+\Big[V_{k}\rho_{\nu},V_{k}^{\dag}\Big]\,,
\label{i.extra}
\end{equation}
where $V_{k}$ are dissipative operators that arises from interaction between the subsystem of interest and the quantum reservoir. So, $V_{k}$ may describe the dissipative effect of the neutrino propagating in quantum gravity space-time \cite{ben1, mavromatos, mavro1, gab, gab1} or in flutuating matter \cite{ben2, bur}. Besides, $V_{k}$ operators act only on the $N$-dimensional $\rho_{\nu}$ space. In the evolution equation (\ref{i}), the first term on right side evolves the quantum state like Liouville equation, while the second term, which depends on $V_{k}$ operators, becomes the evolution non-unitary. The second term is also responsible for introducing many kinds of the dissipative effects ~\cite{lin, len, gor, pet}

We assume that the standard oscillation Hamiltonian is the same for neutrinos as subsystem of interest \cite{workneut4}.  For next generation of long  baseline experiment, like DUNE \cite{lbne}, the matter density is important and the Hamiltonian can be written as      
\begin{equation}
H_{S} = H_{osc}+V_{cc}=diag.\{\tilde{E}_{1},\tilde{E}_{2},\tilde{E}_{3}\}\,,
\label{ii}
\end{equation}
where the $V_{cc}$ is usually the matter potential and on right side of the last equality, we are assuming to be possible to write the Hamiltonian in effective mass basis.

For our aim, it is useful to expand the Eq.~(\ref{i}) in the $SU(3)$ and $SU(2)$ basis matrices. Then, each operator in Eq.~(\ref{i})  can be expanded like $O_{\mu}=a_{\mu}F^{\mu}$, where $F_{\mu}$ is composed by an identity matrix and the Gell-Mann and Pauli matrices for three and two dimension, respectively. Thus, the evolution equation can be rewritten as
\begin{equation}
\frac{d}{dx}\rho_{k}(x)F_{k}=2\epsilon_{ijk}H_{i}\rho_{j}(x)F_{k}+D_{k l}\rho_{l}(x)F_{k}\,,
\label{iii}
\end{equation}
with $D_{\mu 0} =D_{0\nu}=0$ to keep the probability conservation and we change $t \rightarrow x$ as usual for ultra-relativistic approximation. As $\dot{\rho}_{0}(t)=0$, its solution is trivial and it is just $\rho_{0}(t)=1/N$, where N is the number of families. We do not include this component in equation above and besides, the component $a_{0}$ from the $V_{k}$ operators must be null.
 
The $D_{kl}$ matrix in Eq. (\ref{iii}) has many parameters even in two-neutrino approximation. However, if we impose $[H_{S},V_{k}]=0$ the $D_{kl}$ is written as 
\begin{equation}
D_{ll}=-diag\{\Gamma_{21},\Gamma_{21},0,\Gamma_{31},\Gamma_{31},\Gamma_{32},\Gamma_{32},0\} 
\label{iv}
\end{equation}
for three-neutrino approach and
\begin{equation}
D_{ll	}=-diag\{\Gamma_{1},\Gamma_{1},0\} 
\label{v}
\end{equation}
for two-neutrino approximation. We relabeled by $\Gamma_{ij}$ all the compositions of the coefficients $a _{k}$ obtained from expansion of the dissipative terms \cite{workneut3}.  

The dissipators in Eqs. (\ref{iv}) and (\ref{v}) add decoherence effects in the propagation. In the case of the dissipator in Eq. (\ref{iv}), each $\Gamma_{ij}$ parameter may have different values for each quantum interference term and the indices $i$ and $j$ are associated with the quantum decoherence between the families $i$ and $j$. In two families this association is not necessary and the double indices can be replaced by only one indice ~\cite{lis, ben, workneut}. 

The constraint $[H_{S},V_{k}]=0$ imposes energy conservation on subsystem of interest which, in this case, are the neutrinos. However, this constraint is very stringent and physically there is not any guarantee that this must occur, because neutrino energy can fluctuate once it is free to interact with the quantum reservoir. It is important to remember that the energy conservation is always satisfied by the global system \cite{workneut}.

We can define another dissipator to violate the constraint before. It is made replacing the null entries in the main diagonal by new parameters in the dissipators (\ref{iv}) and (\ref{v}). These parameters describe another dissipation effect that is called relaxation effect. 

As the decoherence and relaxation effects depend on propagation distance \cite{workneut, workneut4}, solar and astrophysical neutrinos are only able to put stringent bounds on relaxation effects. In general, the coherence terms are averaged out for neutrinos that come from these sources and hence, it is not possible to have any information about the decoherence effect using our model-independent formalism \cite{ workneut4,kim, pall}. On the other hand, the decoherence effect is only one dissipative effect that terrestrial experiments can limit because the oscillation effect is still important in these cases \cite{ workneut4,kim, pall}.

So, we rewrite only the dissipator in Eq. (\ref{v}) adding the relaxation effect. In this last case, we will show how the dissipation and MSW effects together change the usual behavior of the probability in a particular oscillation channel.

Under the condition $[H,V_{k}]\neq 0$, the dissipator considering two-neutrino approach is written as 
%
%
%
\begin{equation}
D_{km}=-\{\Gamma_{1},\Gamma_{1},\Gamma_{2}\}\,.
\label{vi}
\end{equation}
%

In the particular case of the dissipators in Eqs. (\ref{iv}) and (\ref{v}), the off-diagonal elements must be null due to the constraint $[H,V_{k}]=0$. However, we can ignore all off-diagonal terms in (\ref{vi}) due to complete positivity ~\cite{kra}. As $D_{kl}$ must be positive definite, the diagonal parameters must be larger than the off-diagonal ones. Thus, the dissipator in Eq. (\ref{vi}) is enough to study dissipation effects since the off-diagonal parameters only exist if the main diagonal is filled with the decoherence and  the relaxation parameters.

Complete positivity constrains the elements in main diagonal of the dissipators as well.  The dissipators obtained in Eqs. (\ref{iv}) and (\ref{vi}) have the following constraints:  
\begin{eqnarray}
\begin{array}{ l }
\Gamma_{21}= 2 a_{3}^{2}\geq0,\\
\Gamma_{31}=\frac{1}{2} \left(a_{3}+a_{8}\right)^2\geq0,\\
\Gamma_{32}=\frac{1}{2} \left(a_{3}-a_{8}\right)^2\geq0,
\end{array}
\label{viii}
\end{eqnarray}
and
\begin{equation}
2\Gamma_{1}-\Gamma_{2}\geq 0 \,,
\label{vii}
\end{equation}
respectively. It is important to note that each $\Gamma_{ij}$ may have a specific value.

As we do not assume any microscopic environment model \cite{len}, it is not possible to know which are the energy dependence on the $\Gamma$ parameters \cite{lis}. In this work, we are going to consider $\Gamma$ in eq. (\ref{v}) and (\ref{vi}) as being a constant parameter with energy dimension. In general, a power-law dependence for the $\Gamma$  parameter can be defined as an ansatz \cite{lis, farz} where each power could have relation with a particular physical mechanism \footnote{For example, in Ref. \cite{lis} the power-law was defined as $\Gamma=\gamma_{0}(E/GeV)^{n}$ and with $n=2$, the $\Gamma$ agrees with the typical dimension $ E^{2}/M_{P}$ that is the possible decoherence effects due to the interaction between the matter with the space-time foam \cite{ben1,mavro1}. }.

In the literature there are not any analysis from experiments using the dissipative model for three-neutrino oscillations presented in this section. Although for two-neutrino oscillations there are bounds on decoherence effects ~\cite{lis, fo, workneut3}.  However, considering an experiment like DUNE \cite{lbne}, where the oscillation pattern must consider the three-neutrino families, the constraint in Eq. (\ref{viii}) corroborates that complete analysis of the decoherence effect must be made assuming that $\Gamma_{ij}$ may have different values for each parameter. 

\section{Resonance and dissipation in two families}

In order to observe the new behavior in the situation where neutrinos propagate in density constant matter, we are going to use the formalism for two-neutrino oscillations introduced in the previous section. 

Let us consider the two models that use the dissipators given by in Eq.~(\ref{v}) and~(\ref{vi}). We also consider a hypothetical neutrino source whose oscillation channel would be $\nu_{e}-\nu_{\tau}$. In standard matter effect, $\nu_{e}$ interacts with ordinary matter via charge-current while $\nu_{\tau}$ does not. Even considering this hypothetical source, this oscillation channel is an important component of the oscillation among the three families since the $\theta_{13}$ mixing angle is not null. The exotic behavior that we are going to show also depends on this fact.

We can write the Hamiltonian given in Eq.~(\ref{ii}) in its diagonal form, $H_{S}= diag.\{\tilde{E_{1}},\tilde{E}_{2}\}$, using the effective mass basis. 

The relation between the flavor and effective mass basis is made through the transformation
\begin{equation}
\rho_{m}=U^{\dag}\rho_{f}U\,,
\label{x}
\end{equation}
%
where $U$ is a usual unitary matrix, which depends on the effective mixing angle. The oscillation probabilities can be obtained from 
\begin{equation}
P_{\nu_{\alpha}\nu_{\alpha'}} = Tr[\rho_{\alpha}(x) \rho_{\alpha'}(0)]\,,
\label{xi}
\end{equation}
and taking the quantum dissipator in Eq.~(\ref{vi}), the survival probability is written as   
\begin{equation}
P_{\nu_{\alpha}\nu_{\alpha}} =\frac{1}{2}\bigg[1+e^{-\Gamma_{2} x}\cos^{2}2\tilde{\theta}+ e^{-\Gamma_{1} x}\sin^{2}2\tilde{\theta}\cos\left(\tilde{\Delta} x\right)\bigg]\,,
\label{xii}
\end{equation}
where the transition probability can be obtained from $P_{\nu_{\alpha}\nu_{\alpha'}}=1-P_{\nu_{\alpha}\nu_{\alpha}}$. $\tilde{\Delta}$ is defined as
\begin{equation}
\tilde{\Delta}=\frac{\sqrt{(\Delta m \cos 2\theta-A)^{2}+\Delta m^{2} \sin^{2} 2\theta}}{2 E}\,,
\label{xiii}
\end{equation}
where $E$ is the neutrino energy and the survival probability in Eq. (\ref{xii}) depends on $\sin 2\tilde{\theta}$, which is usually expressed as
\begin{equation}
\sin^{2} 2\tilde{\theta}=\frac{\Delta m^{2} \sin^{2} 2\theta}{(\Delta m \cos 2\theta-A)^{2}+\Delta m^{2} \sin^{2} 2\theta}\,,
\label{xiv}
\end{equation}
where $A=2\sqrt{2}G_{F}n_{e}E$.  

The probability in Eq.~(\ref{xii}) shows the decoherence and relaxation effects that are described by $\Gamma_{1}$ and $\Gamma_{2}$ parameters, respectively. One can obtain the survival probability to the case of the quantum dissipator in Eq.~(\ref{v}) straightforward from Eq.~(\ref{xii}) setting $\Gamma_{2}=0$. The standard oscillation probability is obtained when all the $\Gamma_{i}$ parameters are null. 

We analyze the probability (\ref{xii}) supposing cases allowed for constraints given in Eq. (\ref{vii}). So, we hold the following cases: a-) $\Gamma_{2}=2\Gamma_{1}$, b-)  $\Gamma_{2}=\Gamma_{1}$ and c-) $\Gamma_{2}=0$.

\begin{figure}[!t]
\includegraphics[width= 8.5 cm, height=18.5cm]{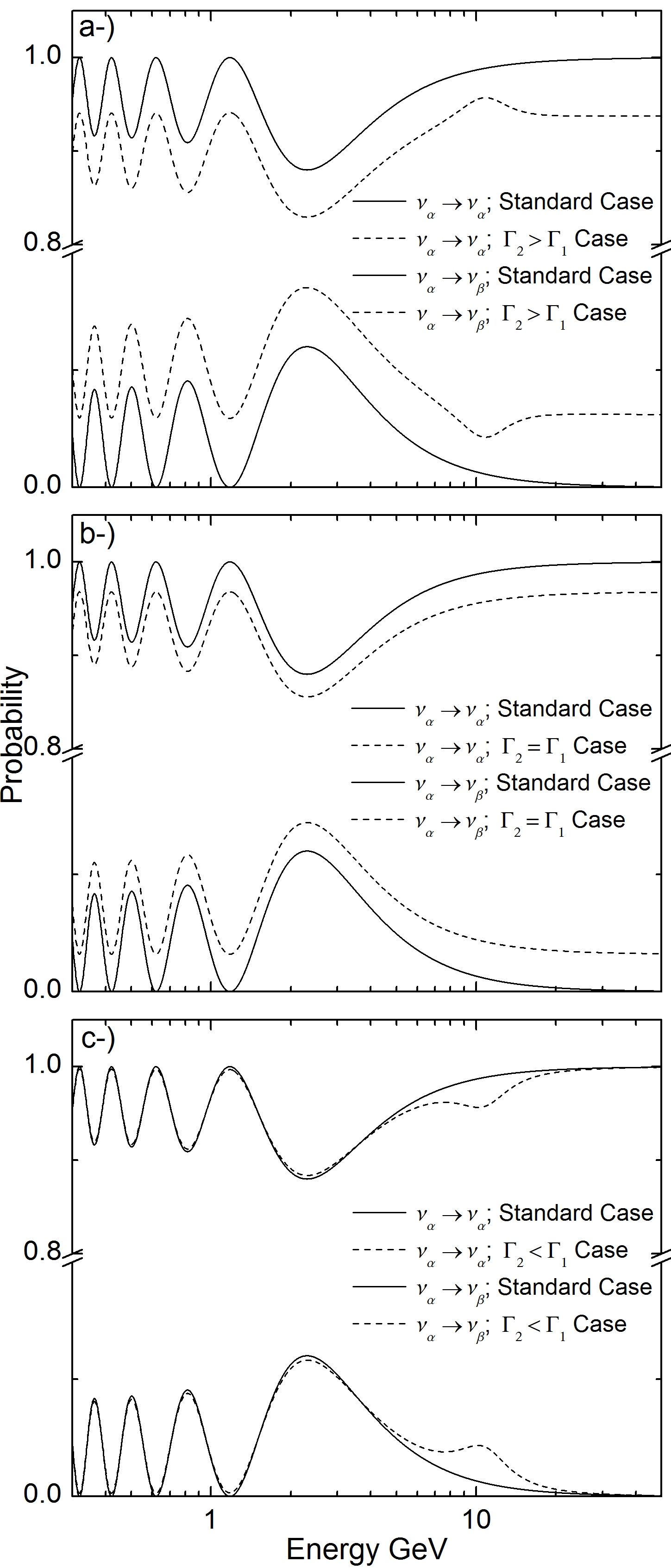}
\caption{These plots are made using $\Gamma_{1}=10^{-23}$ GeV and we vary the $\Gamma_{2}$ parameter. The values used for $\Gamma$ were: in  a-) $\Gamma_{2}=2\Gamma_{1}$, in b-) $\Gamma_{2}=2  \Gamma_{1}$ and in  c-)$\Gamma_{2}=0$.}
\label{fig.i}
\end{figure}

The Fig.~\ref{fig.i} shows the behavior of the oscillation probabilities in all cases analyzed. We consider $\Delta m^{2}= 2.3 \times 10^{-23}$ GeV$^{2}$,  $\theta = 9^{o}$, $L=1300$ km and $\Gamma_{1}=10^{-23}$ GeV that is the current order of magnitude obtained from accelerator and atmospheric experiments in two-neutrino approximation \cite{lis, workneut3}. In this figure, the survival and appearance probabilities for the cases a-), b-) and c-) are represented by upper and lower curves, respectively. 

The cases a-) and b-) in Fig.~\ref{fig.i} shows that relaxation effect becomes the survival probability smaller than the standard one for each energy point, while in the appearance probability occurs the opposite. 

The decoherence effect can be seen through the difference of the oscillation amplitudes between the standard probability and the probability obtained for all dissipative cases. Furthermore, new peaks arise in the cases a-) and c-). These peaks occur around $E\approx 10.52$ GeV that is the resonance region considering the matter density of the Earth crust \cite{kim}. 

From the mathematical point view, these new peaks in cases a-) and c-) can be explained from survival probability in Eq.~(\ref{xii}). Let us write this probability in terms of three functions given by 
\begin{eqnarray}
\begin{array}{ l }
F(\Gamma_{2},x)=\frac{1}{2} +\frac{1}{2}e^{-\Gamma_{2} x};\\
G(\tilde{\theta},\Gamma_{2},x)=-\frac{1}{2} e^{-\Gamma_{2} x}\sin^{2}2\tilde{\theta};\\
H(\tilde{\Delta},\tilde{\theta},\Gamma_{1},x)=\frac{1}{2}e^{-\Gamma_{1} x}\sin^{2}2\tilde{\theta}\cos\big(\tilde{\Delta} x\big),\end{array}
\label{xv}
\end{eqnarray}
such that the survival probability is written as 

\begin{equation}
P_{\nu_{\tau}\nu_{\tau}} =F(\Gamma_{2},x) + G(\tilde{\theta},\Gamma_{2},x)+H(\tilde{\Delta},\tilde{\theta},\Gamma_{1},x) \,.
\label{xvi}
\end{equation}

Due to the relaxation effect described by the damping term in the function $F(\Gamma_{2},x)$, the survival probability assumes smaller values than the standard one. This is a feature of the relaxation effect, where it can change the oscillation probability independently of the oscillation parameters.

\begin{figure*}[!t]
\includegraphics[width= 16.0 cm, height=5.5cm]{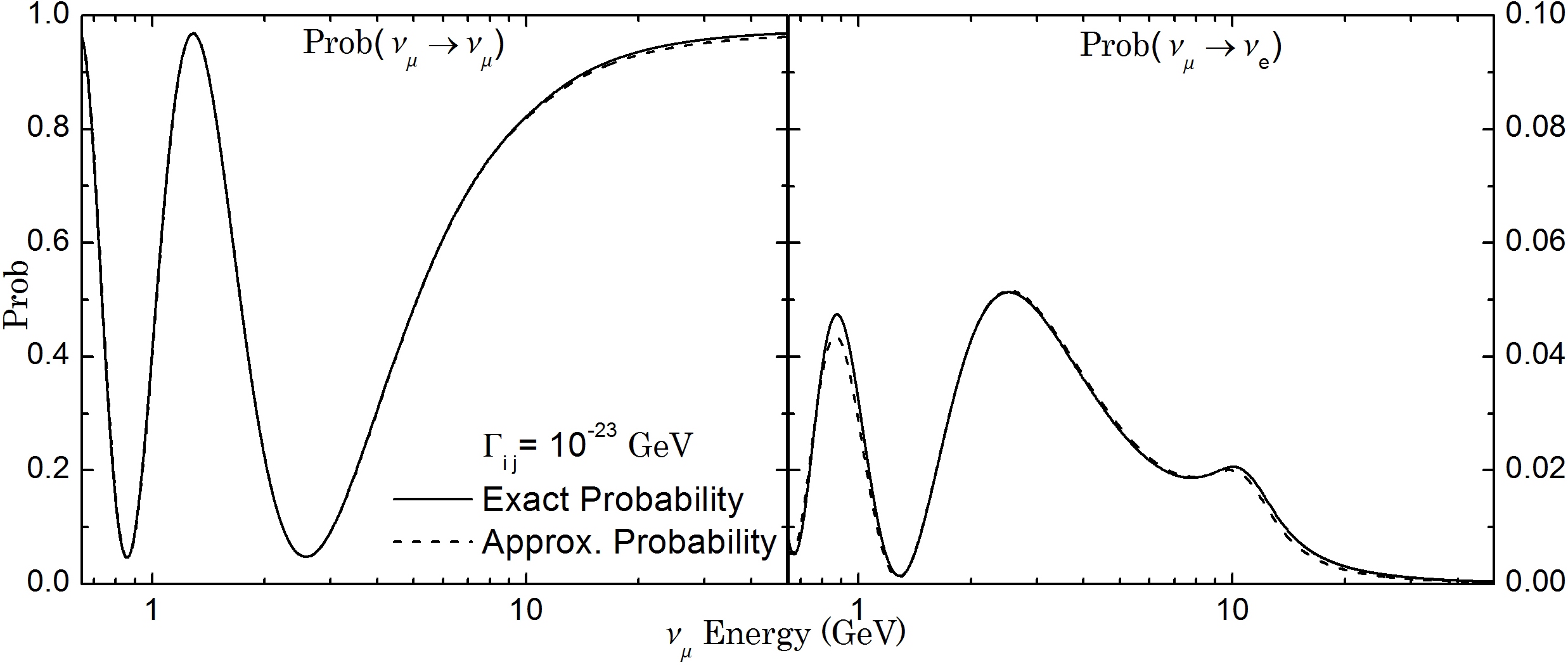}
\includegraphics[width= 16.0 cm, height=5.5cm]{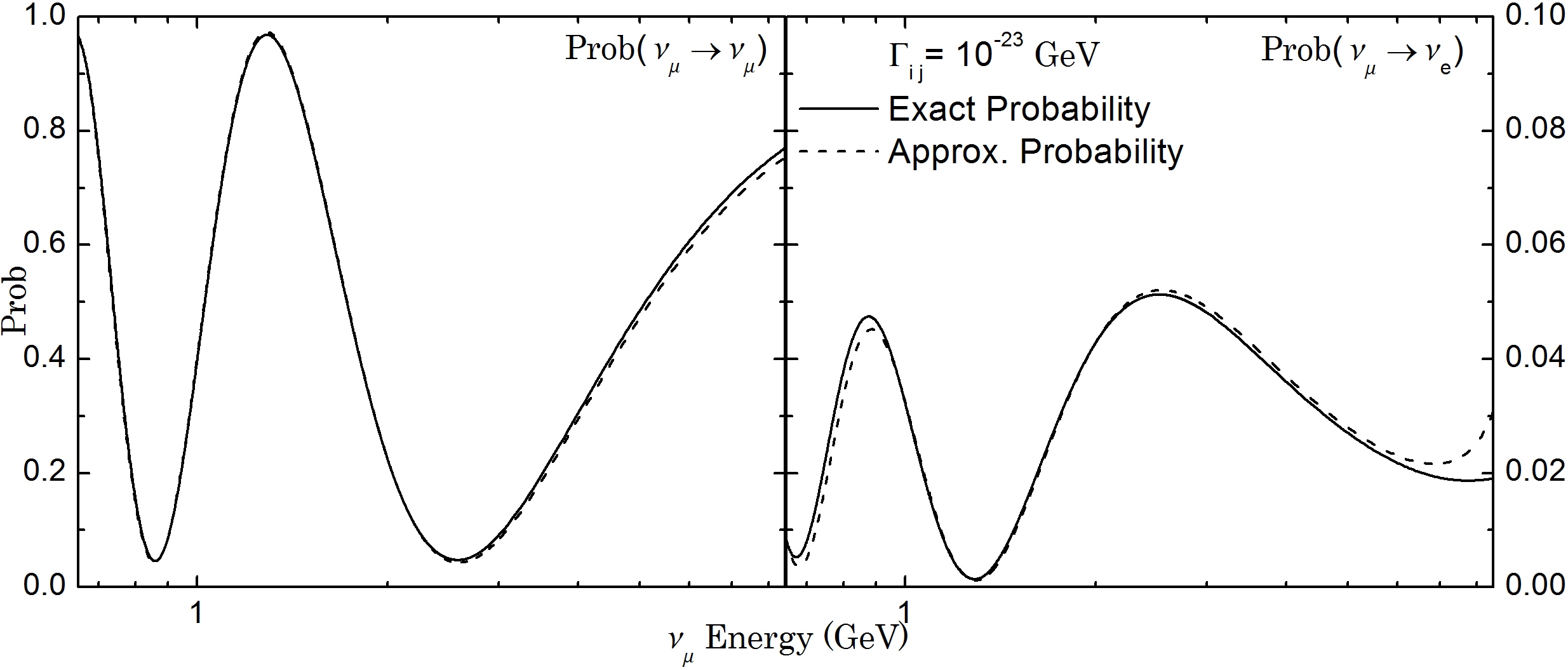}
\caption{The solid lines are for numeric survival (left) and appearance (right) probabilities. The dashed lines are obtained using the probability given in Eq. (\ref{xix}) (top) and at the energy region important for DUNE (bottom), the dashed lines are obtained  using the short approximate probabilities given in Eqs. (\ref{xxiv}) and (\ref{xxviii}). In all cases the decoherence values are $10^{-23}$ GeV.}
\label{figa}
\end{figure*}

At the resonance point ($E_{r}\sim 10.52$ GeV), we have $\cos \tilde{\Delta} x \approx 1$. Then the $H(\tilde{\Delta},\tilde{\theta},\Gamma_{1},x)$ becomes
\begin{equation}
H(\tilde{\Delta},\tilde{\theta},\Gamma_{1},x)|_{E_{r}}\approx h(\tilde{\theta},\Gamma_{1},x)=\frac{1}{2}e^{-\Gamma_{1} x}\sin^{2}2\tilde{\theta}.
\label{xvii}
\end{equation}

In this form, $G(\tilde{\theta},\Gamma_{2},x)$ in Eq.~(\ref{xv}) and $h(\tilde{\Delta},\tilde{\theta},\Gamma_{1},x)$ in Eq.~(\ref{xvii}) are resonance functions with opposite sign and they are different only by $\Gamma_{1}$ and $\Gamma_{2}$ parameters.

When $\Gamma_{1}>\Gamma_{2}$ the resonance peak of the $G(\tilde{\theta},\Gamma_{2},x)$ is smaller than the $h(\tilde{\theta},\Gamma_{1},x)$ and then, we have a peak down (up) in the survival (transition) probability. Taking $\Gamma_{1}=\Gamma_{2}$,  the resonance peak of the $G(\tilde{\theta},\Gamma_{2},x)$ and  $h(\tilde{\theta} ,\Gamma_{1},x)$ have the same amplitude and the sum of them is equal to zero. So, there is not a new peak in this case. On the other hand, when $\Gamma_{1}<\Gamma_{2}$ the resonance peak of  $G(\tilde{\theta},\Gamma_{2},x)$ is larger than the $h(\tilde{\theta},\Gamma_{1},x)$ and then, we have the appearance of the peak up (down) in the survival (transition) probability.

From the phenomenological analysis, the new peaks at the resonance region show that the relation between the MSW and decoherence effects is different from the relation between the MSW and relaxation effects.
 
In the case a-) the relaxation effect is suppressed by the resonance. As we can see, at the resonance region the term $ \cos^{2}(2\tilde{\theta})\sim 0$  in the probability given by Eq. (\ref{xii}). Thus, the new peaks in the probabilities in the case a-) tends to recover the standard behavior at the resonance point. In fact, in Fig. \ref{fig.i} the maximum and minimum values of the peaks do not have the same values of the standard probabilities because $\Gamma_{1}\neq 0$.

The case c-) shows how the decoherence effect still eliminates the quantum interference effect, but in this case, in a very subtle way. The MSW effect changes the amplitude of $\nu_{e}\rightarrow \nu_{\tau}$ with the energy and at the resonance point, the standard oscillation amplitude is maximal, $\sin^{2}(2\tilde{\theta})=1$, but due to the decoherence effect the oscillation amplitude is smaller. At the resonance region the standard survival probability is close to $1$, but the same probability is smaller due to decoherence effect in the case c-). So, in this particular case, the decoherence effect  suppresses weakly the MSW effect increasing the appearance probability at this region.  

So, we can conclude that the relaxation and decoherence act as opposite effects when they are combined with the MSW effect.

Through the Eq. (\ref{xiv}) is clear that the new peaks are not expected for antineutrinos once this equation does not have a resonance behavior in this case. So, the asymmetry neutrino-antineutrino may be changed with the magnitude of the decoherence and relaxation effects.  

As mentioned before, solar and astrophysical neutrinos are responsible by stringent bonds on relaxation effect \cite{workneut4}. Therefore, the terrestrial experiments can be disregarded the cases a-) and b-) and these experiments are able to put bounds only on the decoherence effect and depending on the magnitude of this effect a new peak may exist such as occurred with the case c-).

\section{Analytic Approach for Three-Neutrino Oscillation}

The three-neutrino propagation in constant matter does not have a complete analytic solution and some kind of approximation is necessary in order to obtain the solutions. When it is considered a long baseline neutrino experiment, we can find many approximated solution, as in Refs. \cite{freund, manfred}. 

We are going to use the same condition used by  M. Freund in Ref. \cite{freund} where the solution is obtained from the series expansions up to first order in $\alpha=\Delta m^{2} _{12}/\Delta m^{2} _{31}$ and with this approximation, the author mapped the effective mixing angles that we was reproduced in the Appendix A. Under this condition, we are going to include the decoherence effects in the survival and appearance probabilities.   

The complete Hamiltonian in the flavor basis is expressed as
\begin{equation}
H= \frac{\Delta m^{2}_{31}}{2E} \left[ U
\left(\begin{array}{c c c} 
0 & 0 & 0\\
0 & \alpha & 0\\ 
0 & 0 &  1 \end{array} \right) U^{\dag} +\left(\begin{array}{c c c} 
\hat{A} & 0 & 0\\
0 & 0 & 0\\ 
0 & 0 &  0 \end{array} \right)\right]\,,
\label{xviii}
\end{equation}
where $\hat{A}=2 V E_{\nu}/\Delta m^{2}_{31}$ with $V=\sqrt{2}G_{F}n_{e}$. The $U$ are the usual unitary matrix and we can write the general probability including the decoherence effects as
%
%
%
\begin{figure*}[!t]
\includegraphics[width= 16.0 cm, height=5.5cm]{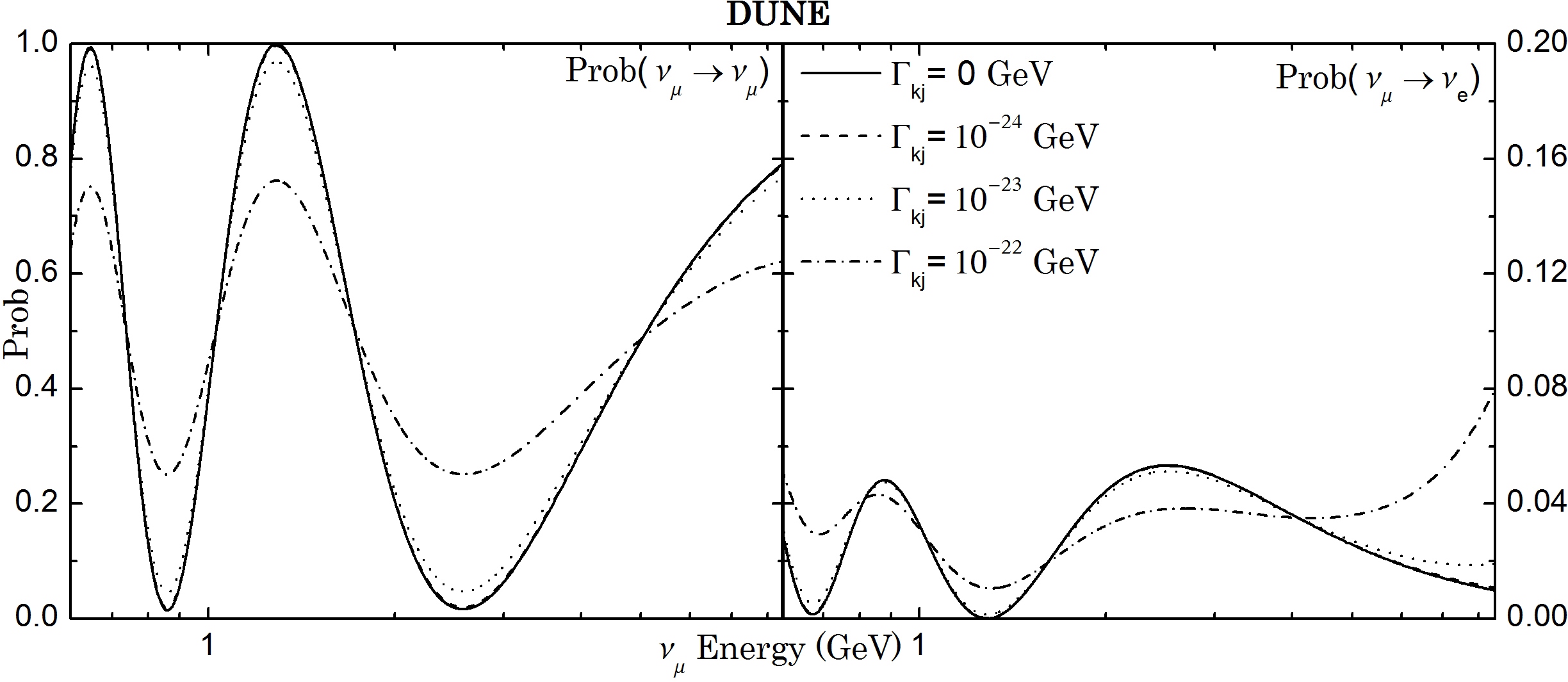}
\caption{Numeric behavior obtained for survival and appearance probabilities considering the DUNE baseline. In these plots was supposed that all the decoherence parameters have the same value.     }
\label{figexp}
\end{figure*}
\begin{equation}
\begin{aligned}
P_{\nu_{\alpha} \nu_{\alpha'}}=& \delta_{\alpha \alpha'}  -2 \sum_{j>k} Re\big(\tilde{U}_{\alpha' j}\tilde{U}^{*}_{\alpha j} \tilde{U}_{\alpha k}\tilde{U}^{*}_{\alpha' k}\big)\\ & +2 \sum_{j>k} Re\big(\tilde{U}_{\alpha' j}\tilde{U}^{*}_{\alpha j} \tilde{U}_{\alpha k}\tilde{U}^{*}_{\alpha' k}\big)e^{- \Gamma_{jk}x}\cos \Delta_{jk} x\\ & -2 \sum_{j>k} Im\big(\tilde{U}_{\alpha' j}\tilde{U}^{*}_{\alpha j} \tilde{U}_{\alpha k}\tilde{U}^{*}_{\alpha' k}\big)e^{- \Gamma_{jk}x}\sin \Delta_{jk} x\,
\end{aligned}
\label{xix}
\end{equation}
where $\tilde{U}_{\alpha i}$ is an element of the approximate effective mixing matrix\footnote{More details can be found in Ref \cite{freund}.} and  all the quartic products in the probability in Eq.~(\ref{xix}) can be obtained from the expressions that we reproduce in Appendix A. The $\Gamma_{jk}$ parameters describe the decoherence effects given in Eq.~(\ref{iv}) and the $\Delta_{jk}$ are the approximate eigenvalues from Hamiltonian (\ref{xviii}). Explicitly, $\Delta_{jk}$ are given by
\begin{eqnarray}
\Delta_{21}&=&\frac{\Delta m_{31}^{2}}{2E}\bigg(\frac{1}{2}(-1-\hat{A}+\hat{C})+\alpha \cos^{2}\theta_{12}\nonumber\\& & - \frac{(1+\hat{C}-\hat{A}\cos 2\theta_{13})\sin^{2}\theta_{12}}{2 \hat{C}} \bigg)	
\label{xx}
\end{eqnarray}
\begin{eqnarray}
\Delta_{32}&=&\frac{\Delta m_{31}^{2}}{2E}\bigg(\frac{1}{2}(1+\hat{A}+\hat{C})-\alpha \cos^{2}\theta_{12}\nonumber\\& & + \frac{(-1+\hat{C}+\hat{A}\cos 2\theta_{13})\sin^{2}\theta_{12}}{2 \hat{C}} \bigg)
\label{xxi}
\end{eqnarray}
\begin{eqnarray}
\Delta_{31}&=&\frac{\Delta m_{31}^{2}}{2E}\bigg(\hat{C}+ \frac{\alpha(1+\hat{A}\cos 2\theta_{13})\sin^{2}\theta_{12}}{\hat{C}} \bigg)
\label{xxii}
\end{eqnarray}
where
\begin{eqnarray}
\hat{C}&=&\sqrt{(\hat{A}-\cos 2\theta_{13})^{2}+\sin^{2} 2\theta_{13}}.
\label{xxiii}
\end{eqnarray}

With the current values for $\alpha$ and $\theta_{13}$ the probability in (\ref{xix}) is an acceptable approximation when they compared with the exact solution \cite{freund}, even when the decoherence effects are taken into account, as it is shown in Fig. \ref{figa}. The standard survival probability obtained from (\ref{xix}) agrees with the exact survival probability, but when the decoherence effects are included, the curves may separate each other from the resonance region depending only on the decoherence magnitude. The standard appearance probability obtained from Eq. (\ref{xix}) and the exact solution have a small discrepancy at low energy region and when the decoherence effects are included a new difference appears at the resonance region.

It is not easy to see how the decoherence effects act on the behavior of the probabilities in Eq. (\ref{xix}). It is possible to obtain short expressions for the probabilities keeping terms proportional to $\alpha^{2}$, $\alpha \sin \theta_{13}$ and $\sin^{2} \theta_{13}$  \cite{freund, manfred}. This procedure allows us to write the survival probability as
\begin{equation}
\begin{aligned}
P_{\nu_{\mu} \nu_{\mu}}\approx& 1-P^{21}_{\nu_{\mu} \nu_{\mu}}-P^{31}_{\nu_{\mu} \nu_{\mu}}-P^{32}_{\nu_{\mu} \nu_{\mu}} \,,
\end{aligned}
\label{xxiv}
\end{equation}
where
\begin{equation}
\begin{aligned}
P^{21}_{\nu_{\mu} \nu_{\mu}}=& (1-e^{\Gamma_{21}x}\cos A\Delta_{21} x)\Big(\frac{\alpha^{2}\cos^{4}\theta_{23}\sin^{2}2\theta_{12}}{2A^{2}}\\ &-\frac{\alpha \cos \delta \cos^{2} \theta_{23}\sin 2\theta_{12} \sin \theta_{13}\sin2\theta_{23}}{(1-A)A}\Big)  \,,
\end{aligned}
\label{xxv}
\end{equation}
\begin{equation}
\begin{aligned}
P^{31}_{\nu_{\mu} \nu_{\mu}}=& (1-e^{\Gamma_{31}x}\cos(1-A)\Delta_{31} x)\Big(\frac{\alpha^{2}\sin^{2}2\theta_{12}\sin^{2}2\theta_{23}}{8A^{2}}\\ &-\frac{\alpha \cos \delta \sin 2\theta_{12} \sin \theta_{13}\sin^{2}\theta_{23}\sin2\theta_{23}}{(1-A)A}\Big)  \,,
\end{aligned}
\label{xxvi}
\end{equation}
and
\begin{equation}
\begin{aligned}
P^{32}_{\nu_{\mu} \nu_{\mu}}=& \Big(\frac{1}{2}\sin^{2}2\theta_{23}-\frac{\alpha^{2}\sin^{2}2\theta_{12}\sin^{2}2\theta_{23}}{8A^{2}}\\ & +\frac{\alpha \cos \delta \sin 2\theta_{12} \sin \theta_{13}\sin2\theta_{23}}{(1-A)A} \times \\& (\sin^{2}\theta_{23}-A^{2}\cos 2\theta_{23})\Big)(1-e^{\Gamma_{32}x}\cos \Delta_{32} x) \,,
\end{aligned}
\label{xxvii}
\end{equation}

and the appearance probability is written as 
\begin{equation}
\begin{aligned}
P_{\nu_{\mu} \nu_{\mu}}\approx& P_{0}+P_{\sin\delta}+P_{\cos\delta}+P_{3} \,,
\end{aligned}
\label{xxviii}
\end{equation}
where 
\begin{equation}
\begin{aligned}
P_{0}=& \frac{\sin^{2}2\theta_{13}\sin^{2}\theta_{23}}{2(A-1)^{2}}\left(1-e^{\Gamma_{31} x}\cos(1-A)\Delta \right)\,,
\end{aligned}
\label{xxix}
\end{equation}
\begin{equation}
\begin{aligned}
P_{3}=&\frac{\alpha^{2}\cos^{2}\theta_{23}\sin^{2}2\theta_{12}}{2A^{2}}\left(1-e^{\Gamma_{21} x}\cos A\Delta \right) \,,
\end{aligned}
\label{xxx}
\end{equation}
\begin{equation}
\begin{aligned}
P_{\cos\delta}=& \frac{\alpha \cos\delta\sin 2\theta_{12}\sin\theta_{13}\sin 2\theta_{23}}{2(A-1)A}
\Big(1-e^{\Gamma_{21} x}\cos A\Delta\\&-e^{\Gamma_{31} x}\cos(1-A)\Delta+e^{\Gamma_{32} x}\cos \Delta \Big)\,,
\end{aligned}
\label{xxxi}
\end{equation}
and
\begin{equation}
\begin{aligned}
P_{\sin\delta}=&\frac{\alpha \sin\delta\sin 2\theta_{12}\sin\theta_{13}\sin 2\theta_{23}}{2(1-A)A}
\Big(-e^{\Gamma_{21} x}\sin A\Delta\\&-e^{\Gamma_{31} x}\sin(1-A)\Delta+e^{\Gamma_{32} x}\sin \Delta \Big) \,,
\end{aligned}
\label{xxxii}
\end{equation}
where $\Delta=\Delta m^{2}_{31}/4 E $.

Disregarding the decoherence parameters, the approximate probabilities in Eqs. (\ref{xxiv}) and (\ref{xxviii}) are continuous function and the apparent divergences for $A\rightarrow 1$ and $A\rightarrow 0$ are canceled by composition of the terms of these probabilities. This same situation was discussed in Ref. \cite{manfred} and besides, without the decorehence parameters in Eq. (\ref{xxviii}) , it is possible to obtain the same expression for the appearance probability found in Ref. \cite{manfred, freund}.

When the decoherence parameters are not null the canceling of the divergences at the resonance region fails even when all decoherence parameters have the same magnitude. Although, if we consider the range energy  important for DUNE experiment, where the use of these probabilities will be interesting, the approximate and the exact probabilities have a similar behavior in most part of the range energy even when the decoherence effect is taken into account, as it is possible to see in Fig. \ref{figa}. The larger difference just occurs for the appearance case at the resonance region depending on the decoherence magnitude.

In concrete case, all calculations for experimental analysis using the probability in Eq. (\ref{xix}) may not have any advantage over the exact approach, even the shorter approximate probability present many terms. However, the probabilities in Eqs. (\ref{xxiv}) and (\ref{xxviii}) are able to show details about the behaviors of the probability in Eq. (\ref{xix}) and the numerical probability. So, we are going to use them to investigate how each decoherence parameter changes the oscillation probabilities.    

\begin{figure*}[!t]
\includegraphics[width= 16.0 cm, height=5.5cm]{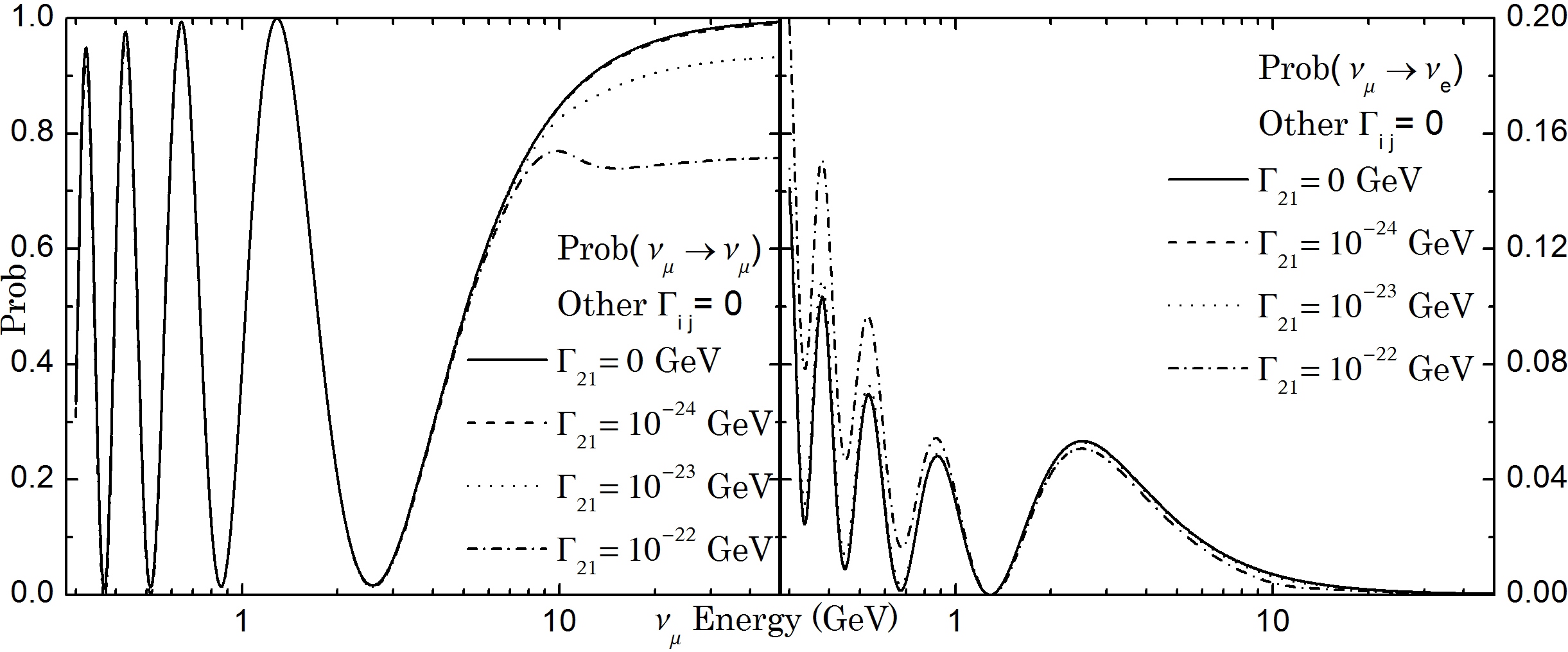}
\includegraphics[width= 16.0 cm, height=5.5cm]{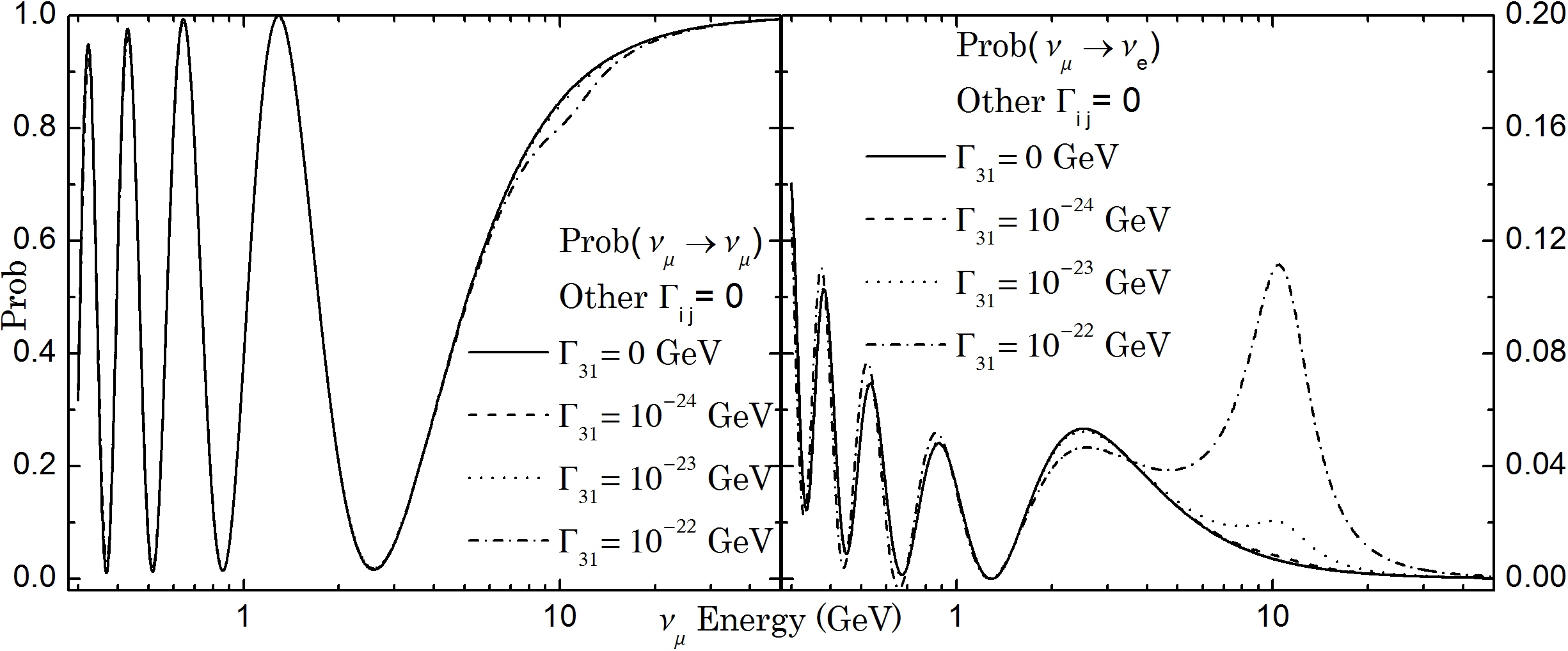}
\includegraphics[width= 16.0 cm, height=5.5cm]{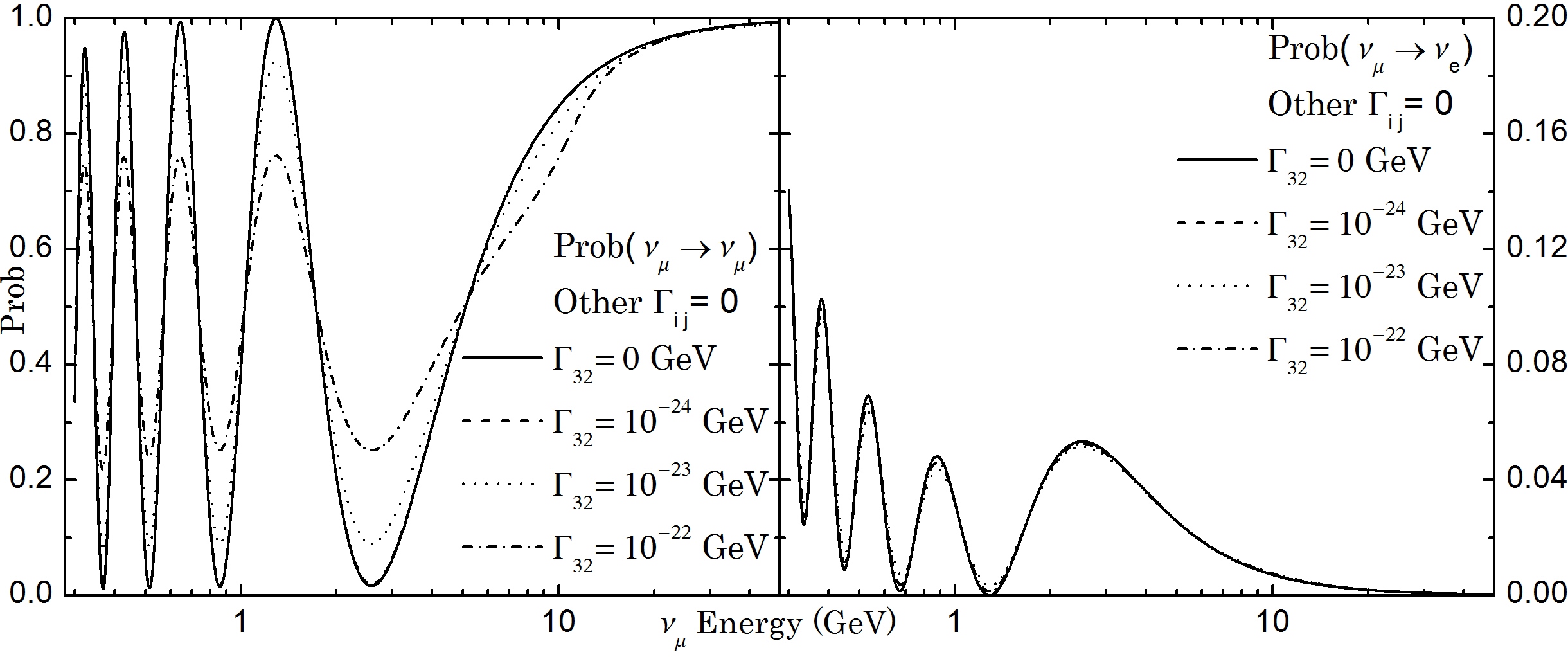}
\caption{This figure shows the behavior expected for each decoherence parameter with different values considering the DUNE baseline. The $\Gamma_{ij}$ describes decoherence effects only between $\nu_{i}-\nu_{j}$ mass states. A new peak due to MSW and the decoherence effect $\Gamma_{31}$  appears at the resonance region.}
\label{fig21}
\end{figure*}

To this end, we are consider the DUNE baseline and use the exact approach to show the behaviors of the probabilities and analytical approach to explain the modifications. It is possible since the Fig. \ref{figa} show the agreement between the analytical and exact approach on the DUNE range energy. For simplicity, we have used the following values for oscillation parameters: $\Delta m^{2}_{12}=8\times 10^{-23}$ GeV, $\Delta  m^{2}_{31}=2.5\times 10^{-23}$ GeV, $\theta _{23}=0.74$, $\theta _{12}=0.58$, $\theta _{13}=0.14$, $\delta =0$ and the usual Earth density~\cite{kim}. However, an important point is that the mixing angles values are close to the current best fit of the literature, but the decoherence effects tend to change the best fit values of the mixing angles.

The Fig. \ref{figexp} shows the curves for the survival and appearance probabilities using the exact approach with different values for the decoherence parameters. For simplicity, we assumed the same value of $\Gamma_{ij}$ for each probability. Besides, these probabilities are limited between $0$ and $1$ as consequence of the complete positivity which is guaranteed by the inequalities in (\ref{vii}).

In the Fig. \ref{fig21} we show the changes due to each term of the decoherence effects. In this case, in order to see as each one these effects alone, the inequalities in the Eq. (\ref{viii}) were not respected.

\subsection{The survival probability}
  
Considering the probability in Eq. (\ref{xxiv}), we can see that the part expressed  in Eq. (\ref{xxvii}) has the dominant oscillation term being proportional only to $\sin^{2} 2\theta_{23}$. Then, $\Gamma_{32}$ would only be responsible for this decrease in the oscillation amplitude. The decoherence parameters $\Gamma_{21}$ and $\Gamma_{31}$ appear in terms that depend on $\alpha^{2} $ or $\alpha \sin\theta_{13}$, so they are subdominant in this oscillation regime. 

As we mentioned before the probability in Eq. (\ref{xxiv}) does not work outside the DUNE energy. From the Fig. \ref{fig21}, we can see the consequences that $\Gamma_{21}$ and $\Gamma_{31}$ bring to the survival probability. In this case, $\Gamma_{31}$ just changes the behavior at the resonance region where the survival probability decreases its value. This is explained thorough the neutrino appearance at this region where it occurs a non-usual appearance due to the MSW effect and the decoherence effect described by  $\Gamma_{31}$. In the same way, $\Gamma_{21}$ also decreases the values for survival probability to high energy. In this case, as $P_{\nu_{\mu}\nu_{\mu}}+ P_{\nu_{\mu}\nu_{e}}+P_{\nu_{\mu}\nu_{\tau}}=1$ and there is not any new neutrino appearance (Fig.~\ref{fig21}), we can just conclude that there is an interesting transition to $ P_{\nu_{\mu}\nu_{\tau}}$ at this region if $\Gamma_{21}$ has enough magnitude.

\subsection{The appearance probability}

In the case of the appearance probability the decoherence parameters may bring exotics peaks at the resonance region. We can use the appearance probability in Eq. (\ref{xxviii}) to explain how the decoherence  effects act on the neutrino behavior.

The $\Gamma_{21}$ parameter tends to decrease weakly the first oscillation peak. The approximate probability does not describe this effect. Only with probability in Eq. (\ref{xix}), that has high order terms, is able to describe this behavior. On the other hand, $\Gamma_{21}$ tends to increase de second peak and the term $P_{3}$ in Eq. (\ref{xxx}) is responsible for this behavior. The $P_{3}$ term do not oscillate at this energy range and its individual value is increased to low energy values, this would contribute to the neutrino appearance.   

The $\Gamma_{31}$ parameter is responsible for the phenomenon that was discussed in the section $III$. However, it is clear that the new peak presented before is outside the range energy of the neutrinos in the DUNE experiment, but it will be possible to bound the $\Gamma_{31}$ parameter through this experiment and depending of the magnitude of this effect, this peak may exist and maybe it can investigated experimentally in other energy configuration that DUNE or another future experiment will may have.

So, considering the current prospect for DUNE and its range energy, the $\Gamma_{31}$ parameter  tends to decrease the appearance probability at the first peak. The important term for this phenomenon is $P_{0}$ in Eq. (\ref{xxix}) because the oscillation amplitude at this energy region is larger than the term $P_{\cos}$ in Eq. (\ref{xxxi}). In addition, when the $P_{cos}$ term has $\Gamma_{31} \neq 0$, it tends to generate a phase difference in relation to $P_{\cos}$ when $\Gamma_{31} =0$, besides the oscillation amplitude of this term is increased when $\Gamma_{31} \neq 0$. Then, the combination between $P_{0}$ and $P_{\cos}$ gives origin to a phase difference in relation to the standard case and the maximum value is slightly larger than the standard case.                 

The phenomenon associated with the $\Gamma_{32}$ parameter are very subtle and high order terms in $\alpha$ are necessary to describe the slight decrease of the amplitude oscillation that is peculiar to the decoherence parameter.  

\begin{figure}[!t]
\includegraphics[width= 8.5 cm, height=18.5cm]{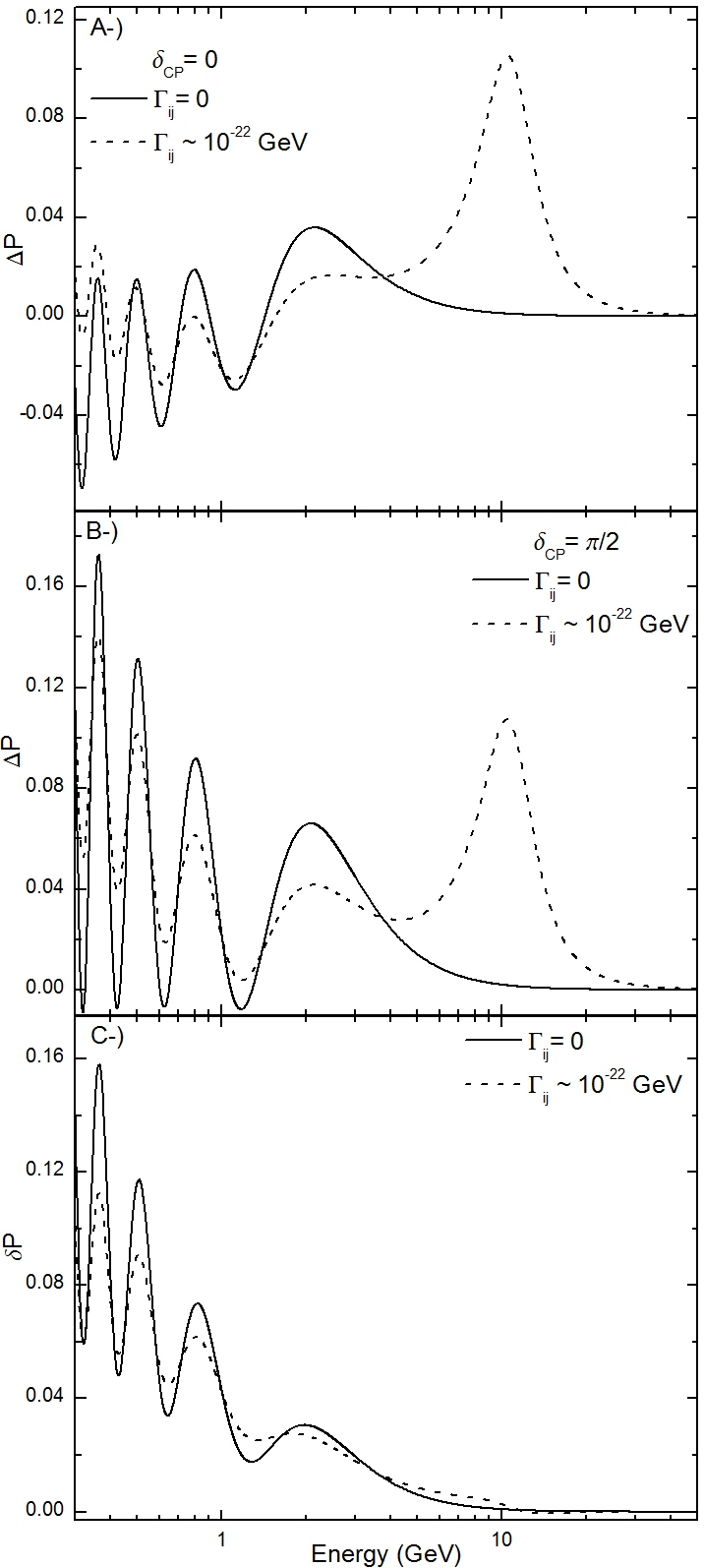}
\caption{Behavior for $\Delta P$ with neutrino CP phase given by $\delta=0$ (top) and $\delta=\pi/2$ (middle). The $\delta P$ is for approximate intrinsic CP-violation (bottom). In all cases the $\Gamma_{ij} =10^{-22}$ GeV.}
\label{fig.cp}
\end{figure}

\section{ CP-Violation}

With the next generation of long baselines experiments there will be the possibilty of to study   CP-violation with sensibility as never before. As each decoherence parameter changes the appearance probability in a specific way, we can expect some change in the CP-violation behavior. In general, the CP-violation can be probed through the quantity called CP-asymmetry which is defined usually as
\begin{equation}
A_{CP}=\frac{P_{\nu_{\mu}\rightarrow\nu_{e}}-P_{\bar{\nu}_{\mu}\rightarrow\bar{\nu}_{e}}}{P_{\nu_{\mu}\rightarrow\nu_{e}}+P_{\bar{\nu}_{\mu}\rightarrow\bar{\nu}_{e}}}\,,
\label{xxxiii}
\end{equation}
and in the case of the dissipative approach, this quantity is also changed due to decoherence effects. However, it is not simple to see the changes thought this definition. So, we are just to use another definition for CP-violation \cite{kim} that is given by 
\begin{equation}
{\Delta P}=P_{\nu_{\mu}\rightarrow\nu_{e}}-P_{\bar{\nu}_{\mu}\rightarrow\bar{\nu}_{e}}\,,
\label{xxxiv}
\end{equation}
where the CP-violation effect is more clear when the decoherence effect is taken into account.

Using the Eq. (\ref{xxxiv}), we plot the Fig. \ref{fig.cp}A  with $\delta=0$ and the Fig. \ref{fig.cp}B with $\delta=\pi/2$. In both cases appear the behaviors of $\Delta P$ with and without the decoherence effects. In Fig. \ref{fig.cp}A the CP-violation is due to the matter potential, usually called of fake CP-violation because the CP phase is equal to zero (or $\delta=\pi$). In Fig. \ref{fig.cp}B the CP-violation is due to CP phase that is $\pi/2$ and also due to matter potential. In both Figs. \ref{fig.cp}A and \ref{fig.cp}B the peak is due to the MSW and $\Gamma_{13}$ effects as we have seen before and it is weakly influenced by $P_{\bar{\nu}_{\mu}\rightarrow\bar{\nu}_{e}}$ that is basically constant at this region. 

In order to see how decoherence effects may change the CP-violation behavior, we can use the Eq. (\ref{xxxiv}). However, through this equation we can only see the action of the decoherence effects on the intrinsic and fake CP-violation as a whole. The matter effect can be lessened through the choice of the baseline, but it is not completely eliminated. In any approximation for neutrino propagation in matter, the matter potential is kept and these two types of CP-violation will be involved. Then, only in approximate way we can see how the decoherence effects change the intrinsic CP-violation. For this end, we can use the following definition \cite{hiroshi}
\begin{equation}
{\delta P}=\Delta P_{\nu_{\mu}\rightarrow\nu_{e}}(\delta=\pi/2)-\Delta P_{\nu_{\mu}\rightarrow\nu_{e}}(\delta=0)\,,
\label{xxxv}
\end{equation}
where we the matter effect is lessened as it is possible to see through the short approximate appearance probability in Eq. (\ref{xxviii}) and so, we can see approximately how the intrinsic CP-violation is changed in Fig. \ref{fig.cp}C. In this figure, the peak at the resonance region disappears and the decoherence effects decrease the amplitude oscillation of the $\delta P$.

\section{Comments and Conclusion}

We have investigated from a phenomenological point view the dissipative effects in neutrino propagation considering long baseline experiments. In our discussion, we supposed a baseline of the DUNE experiment \cite{lbne}. This experiment will have greatest sensibility to many oscillation parameters.  Besides, it will be able to investigate the open questions in neutrinos physics and many non-standard models will be probed from this experiment \cite{lbne}.

Our approach included dissipative effects with a treatment where all quantum interpretations remain as usual. All dissipative effects were obtained from phenomenology arguments based in complete positivity constraint. From this, we introduce the quantum dissipator in Eq. (\ref{iv}) as being the most effective among all that we can obtain for terrestrial experiments. It contains three possible decoherence parameters since the other two possible relaxation parameters are stringently bounded  by solar neutrinos. These parameters are correlated by inequalities in Eq. (\ref{viii}) where at least two decoherence parameters have to be non-null in order to remain the physical evolution.

Due to the new effect introduced by combination between the MSW and the dissipative effects, we made a study in two neutrino approximation to explain how the decoherence and relaxation effects react when the MSW is present. In this case, we have shown that the decoherence and relaxation effects brought a different behaviors to the probabilities. These behaviors are evidenced trough the new up and down peaks at the resonance region.

For three neutrino oscillations was used the usual approximation where one assumes $\Delta m^{2}_{12}<<\Delta m^{2}_{31}$ and we rewrite the important mixing coefficients in Appendix A that was mapped in Ref. \cite{freund}. So, we presented in analytical way two versions for the dissipative model obtained from the dissipator defined in Eq. (\ref{iv}). The first version defined by Eq. \ref{xix}, we used only the coefficients in the Appendix A to calculate all quartic product of the $\tilde{U}_{\alpha k}$ in Eq. (\ref{xix}). So, we can know the approximate behaviors of the survival and appearance probabilities in this case.  These approximate probabilities and numerical solution have similar behaviors in most part of the energy spectrum as we can see in Fig. \ref{figa}.

In the second version, only terms proportional to $\alpha^{2}$, $\alpha \sin \theta_{13}$ and $\sin^{2} \theta_{13}$  are kept in the probabilities. In these case, the addiction of the decoherence parameters become the probabilities divergent at the resonance region, but the Eqs. (\ref{xxiv}) and (\ref{xxviii}) are useful to understand the behaviors of the survival and appearance probabilities in the important range energy of the DUNE experiment.

The Fig. \ref{fig21} presented all decoherence parameters individually showing how each decoherence parameter may change the behavior of the probabilities. The parameters $\Gamma_{12}$ and $\Gamma_{31}$ are important in the appearance case while the $\Gamma_{32}$ is important in survival case. In particular, the dissipative and MSW effect was investigated in detail in two neutrino oscillation where at the resonance region appears new peaks due to action of the both MSW and dissipative effects. In three neutrino oscillation the peak due to decoherence effect appears too, but in this case, it is described by  $\Gamma_{31}$  parameter. 

We finished this work showing the consequences of the decoherence effects on the CP-violation for specific angles where in an approximate way, we can distinguish the fake and intrinsic CP-violation case. As we have seen the decoherence effect  tends to decrease  the amplitude behavior of the CP-violation effects as it is possible to see in Fig. \ref{fig.cp}.

So, we have shown how the decoherence effects may be a non-usual standard phenomena interesting to be investigated in next generation of neutrino experiments. New effects and behavior may be expected and their absences will be due to stringent limits on the decoherence effect.

\begin{acknowledgments}
The author thanks the Gouvea, A. and the Northwestern University where most part of this work was made and he thanks for the support of funding grants 2012/00857-6 and 2013/11651-2 São Paulo Research Foundation (FAPESP).
\end{acknowledgments}


\appendix
\section{ Mixing Angle Functions }
We reproduce here effective mixing angle funtion obtained in Ref.~\cite{freu} that are useful for calculate the quadic products in the probability in~(\ref{xix}). They are as following: 
\begin{eqnarray}
\sin \tilde{\theta}_{13}&=&\frac{\sin 2\theta_{13}}{\sqrt{2\hat{C}(\hat{C}-\hat{A}+\cos 2\theta_{13})}}\nonumber \\ &&+\frac{\alpha \hat{A} \sin^{2}\theta_{12}\sin^{2} 2\theta_{13}}{2 \hat{C}^{2}\sqrt{2\hat{C}(\hat{C}+\hat{A}-\cos 2\theta_{13})}}
\label{A.i}
\end{eqnarray}
\begin{equation}
\sin \tilde{\theta}_{12}=\frac{ \alpha \hat{C}\sin 2\theta_{12}}{ |\hat{A}|\cos\theta_{13}\sqrt{2\hat{C}(\hat{C}-\hat{A}+\cos 2\theta_{13})}}
\label{A.ii}
\end{equation}
\begin{equation}
\sin \tilde{\theta}_{23}=\frac{\alpha \hat{A}\cos \delta  \sin 2\theta_{12}\sin \theta_{13} \cos\theta_{23}}{-1-\hat{C}+\hat{A}\cos 2 \theta_{13}}+\sin \theta_{23}
\label{A.iii}
\end{equation}
\begin{equation}
\sin \tilde{\delta}=\sin \delta\left(-1+\frac{2 \alpha \hat{A}\sin 2\theta_{12}\sin \theta_{13}\cos\delta}{\tan 2\theta_{23} (1+\hat{C}-\hat{A}\cos 2 \theta_{13})}\right)
\label{A.iv}
\end{equation}
\begin{eqnarray}
\sin^{2} 2\tilde{\theta}_{13}&=&\frac{2 \alpha \hat{A} (-\hat{A}+\cos 2\theta_{13})\sin^{2}\theta_{12}\sin^{2}2\theta_{13}}{ \hat{C}^{4}}\nonumber \\ &&+\frac{\sin^{2} 2\theta_{13}}{\hat{C}^{2}}
\label{A.v}
\end{eqnarray}
\begin{equation}
\sin 2 \tilde{\theta}_{12}=2\sin \tilde{\theta}_{12}
\label{A.vi}
\end{equation}
\begin{equation}
\sin 2\tilde{\theta}_{23}=\frac{2 \alpha \hat{A}\cos \delta  \sin 2\theta_{12}\sin \theta_{13} \cos\theta_{23}}{-1-\hat{C}+\hat{A}\cos 2 \theta_{13}}+\sin 2\theta_{23}
\label{A.vii}
\end{equation}

where all there expression consider only first order in $\alpha$. More simplification can be done if we disregard the subleading terms that come from terms that depend on $\alpha \times \theta _{13}$.




\end{document}